\documentclass[sigconf]{acmart}

\AtBeginDocument{%
  \providecommand\BibTeX{{%
    \normalfont B\kern-0.5em{\scshape i\kern-0.25em b}\kern-0.8em\TeX}}}

\copyrightyear{2021} 
\acmYear{2021} 
\setcopyright{acmlicensed}\acmConference[ICAIF'21]{2nd ACM International Conference on AI in Finance}{November 3--5, 2021}{Virtual Event, USA}
\acmBooktitle{2nd ACM International Conference on AI in Finance (ICAIF'21), November 3--5, 2021, Virtual Event, USA}
\acmPrice{15.00}
\acmDOI{10.1145/3490354.3494375}
\acmISBN{978-1-4503-9148-1/21/11}

\begin{document}

\title{Risk and return prediction for pricing portfolios of non-performing consumer credit}

\author{Siyi Wang}
\authornote{Equal contribution.}
\affiliation{%
  \institution{School of Data Science\\
  City University of Hong Kong}
  \country{}
}
\email{swang348-c@my.cityu.edu.hk}

\author{Xing Yan}
\authornotemark[1]
\affiliation{%
  \institution{Institute of Statistics and Big Data\\
  Renmin University of China}
  \country{}
}
\email{xingyan@ruc.edu.cn}

\author{Bangqi Zheng}
\affiliation{%
  \institution{JD Digits}
  \country{}
}
\email{zhengbangqi@jd.com}

\author{Hu Wang}
\affiliation{%
  \institution{JD Digits}
  \country{}
}
\email{wanghu5@jd.com}

\author{Wangli Xu}
\affiliation{%
  \institution{Center for Applied Statistics and School of Statistics\\
  Renmin University of China}
  \country{}
}
\email{wlxu@ruc.edu.cn}

\author{Nanbo Peng}
\affiliation{%
  \institution{JD Digits}
  \country{}
}
\email{pengnanbo@jd.com}

\author{Qi Wu}
\authornote{Corresponding author.}
\affiliation{%
  \institution{School of Data Science\\
  City University of Hong Kong}
  \country{}
}
\email{qiwu55@cityu.edu.hk}

\renewcommand{\shortauthors}{Wang and Yan, et al.}

\begin{abstract}
We design a system for risk-analyzing and pricing portfolios of non-performing consumer credit loans. The rapid development of credit lending business for consumers heightens the need for trading portfolios formed by overdue loans as a manner of risk transferring. However, the problem is nontrivial technically and related research is absent. We tackle the challenge by building a bottom-up architecture, in which we model the distribution of every single loan's repayment rate, followed by modeling the distribution of the portfolio's overall repayment rate. To address the technical issues encountered, we adopt the approaches of simultaneous quantile regression, R-copula, and Gaussian one-factor copula model. To our best knowledge, this is the first study that successfully adopts a bottom-up system for analyzing credit portfolio risks of consumer loans. We conduct experiments on a vast amount of data and prove that our methodology can be applied successfully in real business tasks.
\end{abstract}

%



\maketitle

\section{Introduction}

In some countries, the recent rapid development of consumer credit lending business which lies in the center of FinTech industry \cite{claessens2018fintech}, has changed a large population's financial lifestyle and consumption patterns. 
It enables credit lending to those non-standard customers (to traditional commercial banks), and more flexible credit services are provided than in traditional credit card business. The credit providers are usually some big online platforms that have advantages in technology and data. They adopt many data-driven AI techniques to manage the whole workflow of the business.

While it has the potential of benefiting consumption promotion and activating small and medium-sized enterprises' economic activities \cite{hau2019fintech}, it faces many challenges when running this business, among which the inherent risk is the core \cite{ryu2018understanding}.
The most important challenge and technical problem in this business is the management of risks throughout the entire life stages of all loans, using the accumulated large amounts of data and the most advanced AI techniques. 
In fact, there have been lots of industry practices such as credit scorecards for managing the default-related risks, but some issues still remain unsolved and need the design of new tools.

While most existing research works and industry practices concentrate on the default risk of every single loan, an unresolved but extremely important problem is the risk of a large pool of loans, or a loan portfolio. Analyzing such type of risk is crucial and necessary in both the pricing and the regular risk management of non-performing loan portfolios, i.e., the loans that have been overdue. 
Such portfolios can form asset-backed securities, the trading of which can be used as a manner of risk transferring. This is in great need when many lending providers have accumulated large amounts of risky loans and want to trade them for various purposes.

Overdue loans are sensitive to economic status, or they share some common underlying economic risk factor. Only modeling well the risk of every single loan is far away from determining the overall risk of a loan portfolio. Thus, developing a model system that can incorporate both the modeling of single risk and the capturing of overall risk is essential for non-performing loan portfolios. Mathematically, this needs us to determine the marginal distributions of the single risk-related quantities as well as the dependence structure among them, which is the goal of this paper.

In this paper, we consider modeling both the repayment rate of each overdue loan in a portfolio, and the overall repayment rate of the whole portfolio, in the distribution sense. When a loan is not overdue, generally one will model its default probability. Otherwise, the overdue loan is treated as a defaulted asset and its repayment rate (or recovery rate, a term in the literature of corporate bonds) is considered to be modeled by us, which is a mixed variable of discrete and continuous. Moreover, in finance, the term risk refers to the uncertainty of a quantity related to one's profit and loss, and is measured by some properties of the quantity's distribution.
Hence for the need of risk pricing, we will focus on the prediction of a variable's distribution, which is very different from the single-value prediction in traditional machine learning research.

\subsection{Literature Review}

There has been a large body of literature focusing on the problem of credit scoring of the single customer or loan. Recent papers include \cite{ignatius2018fuzzy}, \cite{fang2019new}, \cite{beque2017approaches}, \cite{babaev2019rnn}, \cite{kang2019novel}, \cite{li2019transfer}, etc., or the comprehensive review \cite{thomas2017credit}. 
Techniques for credit scoring include logistic regression \cite{jung2015rebuild}, support vector machines \cite{carrizosa2017clustering}, \cite{harris2015credit}, 
artificial neural networks \cite{pacelli2011artificial}, \cite{malhotra2003evaluating}, \cite{abdou2008neural}, \cite{babaev2019rnn}, and so on.
Given the existing studies today, few of them take care of the non-performing credit, i.e., the overdue loans.

Moreover, from another point, very few papers considered the default relationships among a group of loans. What we can find is in \cite{rosch2004forecasting}, authors analyzed the retail portfolio credit risk with lagged macroeconomic risk drivers.
\cite{madeira2019measuring} used simulation method to study the default conditions of consumer loans and concluded that consumer loans have a high covariance beta. \cite{cheng2020contagious} studied default contagious chains in the guarantee network in order to identify potential systemic financial risk.

In fact, studies on credit portfolio risk have existed for a long time, but the focus is on corporate bonds and the pricing of CDO (collateralized debt obligations). The examples include \cite{li2000default}, \cite{laurent2005basket}, \cite{wang2009pricing}, \cite{hull2004valuation}, \cite{peng2009default}, \cite{giesecke2010exact}, and the book \cite{o2011modelling}.
The key technical problem of these works is similar to one of ours in this paper: capturing the dependence structure or correlations among loans.
They all adopted a factor copula model to capture the correlation structure among the bonds in a basket. But different from the situation we face, traders in financial markets can imply the parameters of the model from the market CDO quotes using the pricing formula, while we must infer the model parameters from the data collected because there are no active trades happened or there is no such market.
Given the data of overdue loans with unique characteristics such as extremely heterogeneous and ultra high-dimensional, the technical problem we need to solve is a challenging task with no attempts before as we know.

\subsection{Contributions}

To summarize, the contributions of our paper are three-folds:

1) We build a bottom-up architecture for estimating and predicting non-performing consumer credit portfolios' repayment risk with overdue loan data. The data is of mixed type, extremely heterogeneous, and ultra high-dimensional.
We tackle the challenge that few researchers have attempted before in the existing literature.

2) Specifically, we predict the marginal distribution of each single loan's repayment with simultaneous quantile regression, deal with the mixed repayment rate issue with R-copula, model the correlation structure of loans via one-factor Gaussian copula, 
and finally use Monte Carlo simulation to obtain the distribution of the portfolio's overall repayment rate.

3) We show in the experiments that our system can be applied successfully in real business tasks, including the pricing and the risk management of overdue loan portfolios. It may meet the needs of rapidly growing business of risk transferring.

\section{PRELIMINARIES}

\subsection{Quantile Regression}
\label{sec:pre.quantile.regression}

Quantile regression \cite{koenker2001quantile} \cite{meinshausen2006quantile} is a powerful tool that enables us to look at the slices of the conditional distribution $p(y|x)$ without any assumptions on the global distribution. 
Recently there have been very successful adoptions of multi-level quantile regression or its extension to financial forecasting problems, such as tail risk forecasting in financial markets \cite{wu2019capturing} and loss prediction of defaulted bank loans \cite{kellner2021opening}.
We utilize multi-level quantile regression to approximate the full marginal distribution of each loan's repayment rate, as shown in later sections.

Suppose a random variable $Y$ has cumulative distribution function $F_Y$, the $\tau$-quantile of $Y$ is defined as $Q_Y(\tau)=F_Y^{-1}(\tau), \tau\in (0,1)$.
If we further assume that for a fixed $\tau$, the conditional quantile of $Y$ given another variable $X$ is a parameterized function: $Q_Y(\tau|X=x) = f_\theta(x)$, then the function $f_\theta$ can be estimated using quantile regression by minimizing the following expected loss:
\begin{equation}
       \min_{\theta} E[L_\tau(Y,f_\theta(X))],
\end{equation}
where the quantile loss function $L_\tau$ is defined as:
\begin{equation}
        L_\tau(y,q)=
        \begin{cases}
        \quad\quad   \tau|y-q|,\quad &y> q,\\
        (1-\tau)|y-q|,\quad &y\le q.
        \end{cases}
\end{equation}
Given a dataset $\{(x_i, y_i)\}_{i = 1}^{N}$, the estimation of $f_\theta$ is done through minimizing the empirical loss: $\min_{\theta} \frac{1}{N}\sum_{i=1}^{N} L_\tau(y_i,f_\theta(x_i))$.
After solving this minimization problem to obtain the optimal $\theta^*$, the conditional quantile of $Y$ given $X=x$ can be predicted by $f_{\theta^*}(x)$.

The above procedure is for a fixed $\tau$ only, or, for the conditional quantile of $Y$ at a fixed probability level only. This local information of the conditional distribution $p(y|x)$ is not sufficient when we aim to have an estimation of the full distribution. A direct solution is to estimate different quantiles for different probability levels $\tau_1 < \tau_2 < \cdots< \tau_K$ simultaneously and adopt an interpolation method to obtain the full distribution. Generally in simultaneous quantile regression, a $K$-dimensional parameterized mapping $f_{\theta}(x)$ is specified and its $k$-th element $[f_{\theta}(x)]_k$ represents the conditional $\tau_k$-quantile of $Y$. Then, the optimal $\theta^{*}$ is found through solving the following optimization problem:
\begin{equation}
   \min_{\theta} \frac{1}{N}\frac{1}{K}\sum_{i=1}^{N}\sum_{k=1}^{K} L_{\tau_k}(y_i , [f_{\theta}(x_i)]_k).
\end{equation}

However, after this, we may find that the resulting quantiles $f_{\theta^*}(x)$ may cross, i.e., $[f_{\theta^*}(x)]_j>[f_{\theta^*}(x)]_k$ for some $x$, $j$, and $k$ satisfying $j<k$.
This violates the basic principle that a cumulative distribution function and its inverse, i.e., the quantile function, should be monotonically  increasing. This may occur because we do not impose any restrictions on the elements of $f_{\theta}(x)$ in the optimization.
An easy solution is rearranging the order of $[f_{\theta^*}(x)]_1,\dots,[f_{\theta^*}(x)]_K$ such that they are monotonically  increasing \cite{chernozhukov2010quantile}. This is how we tackle this issue in this paper, although there are also other solutions \cite{takeuchi2006nonparametric} \cite{bondell2010noncrossing} which are more sophisticated.

\subsection{Factor Copula Model}

Probability integral transform is a way to convert any continuous random variable following a given distribution to a uniformly distributed random variable. This enables the departure of marginal distribution modeling and dependence modeling among continuous random variables $Y_1,Y_2,\dots,Y_n$, which is called the copula approach. Specifically, suppose $Y_i$ has cumulative function $F_i$, then $U_i := F_i(Y_i) \sim U(0, 1)$. One can define a copula function $C(u_1,u_2,\dots,u_n)$ to describe the joint cumulative function of $U_1,U_2,\dots,U_n$ instead of modeling  $Y_1,Y_2,\dots,Y_n$ directly.
This is particularly useful when every $Y_i$ has different and complex marginals.

In high-dimensional cases, it is very often that all $U_i$ are further converted to some commonly used random variables such as Gaussian or Student's $t$: $V_i : = G^{-1}(U_i)$, where $G$ is the cumulative function of the new random variable specified. Taken Gaussian as an example, now $V_i$ is standard normally distributed and thus we can define dependence structure among $V_1,\dots,V_n$ easily by a full correlation matrix. However, it is more practical to assume a one-factor structure \cite{cousin2008overview} among them in high-dimensional cases:
\begin{equation}
  V_{i}= \beta_i M+\sqrt{1-\beta_i^2} Z_{i},
\end{equation}
where both $M$ and $Z_i$ are assumed to be standard normal variables.
$M$ is a common factor shared by all $V_i$, and $Z_{i}$ is called the idiosyncratic variable. $\beta_i$ measures the sensitivity of $V_i$ to $M$, thus incurs dependence between $V_i$ and $V_j$ when $\beta_i>0$, $\beta_j>0$, $i\ne j$. The correlation between $V_i$ and $V_j$ is $\beta_i \beta_j$.


\subsection{R-copula}

In most cases the random variable we model is continuous, thus the ordinary copula approach can be applied. However, in some special cases, the variable is the mixed type of discrete and continuous. The repayment rate of non-performing consumer credit is a typical example and is the focus of this paper. Loan borrowers may repay zero amount with a large probability or repay a continuous amount between zero and the sum of principal and interest.
Now the probability integral transform cannot be applied directly.
We address this issue with the so-called R-copula approach.

If the cumulative function $F$ of the variable $Y$ is not continuous at $y_0$, there must be $F_+(y_0)-F_-(y_0)>0$. Thus the probability integral transform $F(Y)$ will not take value in the interval $(F_-(y_0),F_+(y_0))$. $F(Y)$ is not standard uniform anymore.
To fix it, R-copula \cite{faugeras2012probabilistic}, \cite{ruschendorf2009distributional} proposed to add another uniformly distributed variable to fill up this interval when $y=y_0$:  $F_-(y_0)+U'(F_+(y_0)-F_-(y_0))$, where $U'\sim U(0, 1)$ is independent. The formal definition of this transform called random probability integral transform is
\begin{equation}
    U:=F_-(Y)+U'(F_+(Y)-F_-(Y)).
\end{equation}
When $Y$ takes value at a continuous point, $F_+(Y)=F_-(Y)$ and $U=F(Y)$ as usual. At a discontinuous point, the standard uniformly distributed $U'$ will fill up the gap between $F_-(Y)$ and $F_+(Y)$.
When there are only finite discontinuous points, it is not hard to prove that $U$ follows a standard uniform distribution. Moreover, the inverse transform from $U$ to $Y$ exists and is unique, making this method useful in practice. When we aim to model the dependence among $Y_1,\dots,Y_n$, the new variables $U_1,\dots,U_n$ can be constructed through this transform, and the Gaussian copula approach can be applied as usual. However, this method adds additional randomness caused by $U'$ into $U$, which needs careful treatment by us in real applications.

\section{The method}

\begin{figure*}[t]
  \centering
  \includegraphics[width=0.8\linewidth]{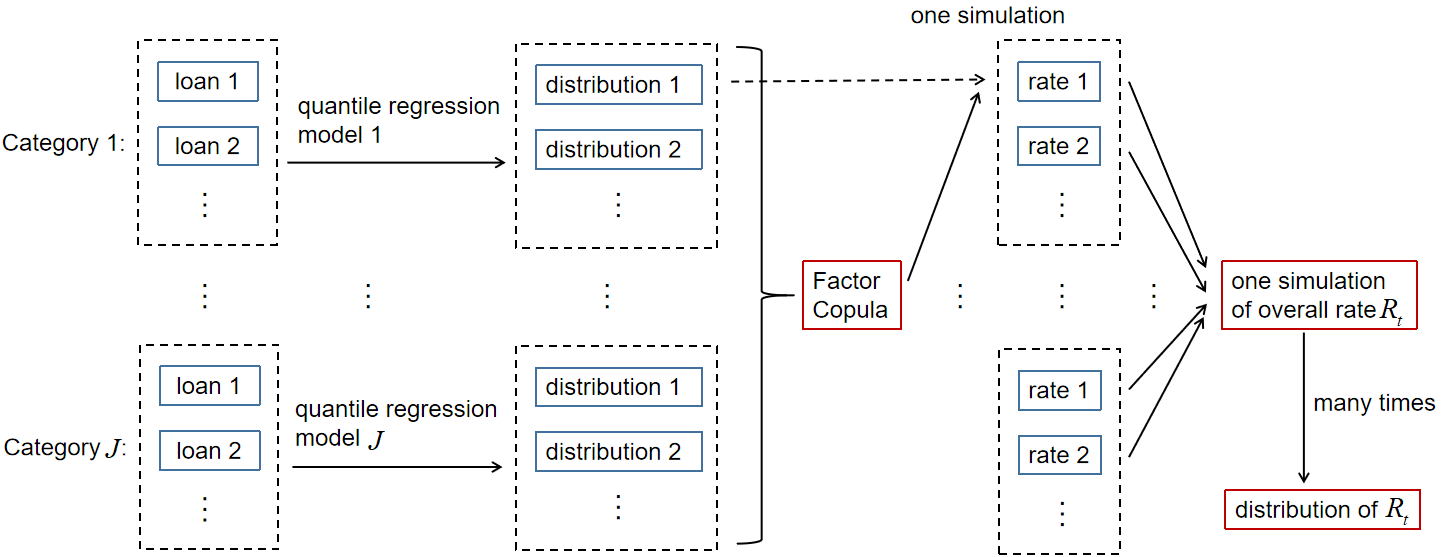}
  \vspace{-1em}
  \caption{The workflow of our system. This is for a fixed time $t$. The model training needs the collection of data over time.}
\label{fig:workflow}
\end{figure*}

In this section, we first describe the problem setup, and then elaborate on the architecture of the method we propose. Our final goal is to predict the distribution of a non-performing consumer credit portfolio's overall repayment rate, as a further input of a pricing 
formula for risky assets.
Our method architecture includes the quantile regression for marginal repayment rate prediction of individual loans, the R-copula approach for the mixed discrete and continuous rates, and the factor copula model for dependence modeling among loans.
After fitting all unknown parameters, we simulate many repayment rates for all loans from our method, and further approximate the distribution of the portfolio's overall repayment rate. The entire workflow is also illustrated in Figure \ref{fig:workflow}.

\subsection{Problem Setup}

Suppose at time $t$, we have $N_t$ loans in the pool that have defaulted. Typically $t$ denotes at which month we are inspecting. Furthermore, these $N_t$ loans are divided into $J$ categories according to their overdue stages, e.g., 30+ days, 60+ days, etc. The $j$-th category has $N_{jt}$ loans included and their repayment rates in a fixed future time horizon are denoted as $R_{ijt}$, where $i\in I_{jt}$ and $I_{jt}$ is the corresponding index set. The repayment rate is defined as
\begin{equation}\label{eqn:repayment rate single}
R_{ijt} = \frac{\text{actual repayment amount}}{\text{expected repayment amount}}.
\end{equation}
We denote the expected repayment amount (generally it is the sum of principal and interest) in the fixed future time horizon as $P_{ijt}$. So $R_{ijt}P_{ijt}$ is the actual repayment amount.
Typically this future time horizon can be chosen to be one year for the need of pricing.

For this portfolio of the $N_t$ loans, the overall repayment rate is 
\begin{equation}\label{eqn:definition of R_t}
R_t = \frac{ \sum_{j=1}^{J}\sum_{i\in I_{jt}} R_{ijt}P_{ijt} }
{ \sum_{j=1}^{J}\sum_{i\in I_{jt}} P_{ijt} }.
\end{equation}
Our goal is to estimate or predict the probabilistic distribution of $R_t$. To achieve this goal, the joint distribution of all $R_{ijt}$ for $j=1,\dots,J$ and $i\in I_{jt}$ should be modeled and estimated. This includes the marginal distribution of every $R_{ijt}$ and the dependence structure among them. These rates are not independent, because the credit status of a large population may be driven by some common underlying risk factors, such as the business cycle or the exogenous shock like COVID-19. In no situation this task will be easy due to the heterogeneity of every loan borrower's credit behavior and the high-dimensional property when considering a huge-size portfolio. We tackle this challenge with the proposed method containing the two core parts described in Section \ref{sec:marginal prediction} and \ref{sec:dependence modeling} respectively.

\subsection{Marginal Distribution Prediction}
\label{sec:marginal prediction}

The first component of our proposed method is predicting the marginal distribution of $R_{ijt}$, whose cumulative function is denoted as $F_{ijt}$. To improve the model flexibility, we do not assume a parametric form for $F_{ijt}$, but instead approximate it non-parametrically by the quantiles of a lot of probability levels spreading all over $(0,1)$. Suppose the probability levels we choose are $\tau_1<\cdots<\tau_K$, the $\tau_k$-quantile of $F_{ijt}$ is denoted as $q_{ijt}^k$. Furthermore, suppose we have the feature vector $x_{ijt}$ for each loan, we use the following model to predict the quantiles of $F_{ijt}$:
\begin{equation}\label{eqn:quantile regression function chosen}
[q_{ijt}^1,\dots,q_{ijt}^K]^\top = \Psi \left( x_{ijt} \mid \Theta_j \right),
\end{equation}
where $\Psi(x|\Theta)$ is a neural network-based model with the parameter set $\Theta$. The subscript $j$ in $\Theta_j$ means we set different models for different categories of loans. $\Psi(x|\Theta)$ is specified as following:
\begin{equation}\label{eqn:choice of psi}
\Psi(x|\Theta) = \text{clip}\left( \text{linear} ( \text{NN}_L (x) ) \right),
\end{equation}
where $\text{NN}_L$ is a $L$-layer ordinary neural network with the sigmoid activation function. The output of $\text{NN}_L$ is then fed to a linear layer and the following clip operator which makes the final output be between the possible minimum and maximum values of $R_{ijt}$.

To train the model for every category, or to learn the parameter set $\Theta_j$ for a fixed $j$, the observations $r_{ijt}$ of $R_{ijt}$ across different $t$ and different $i\in I_{jt}$ are collected to solve the following quantile loss minimization problem:
\begin{equation}
\min_{\Theta_j} \frac{1}{\sum_t N_{jt}} \frac{1}{K} \sum_t \sum_{i\in I_{jt}} \sum_{k=1}^{K} L_{\tau_k} \left( r_{ijt}, [\Psi(x_{ijt}|\Theta_j)]_k \right),
\end{equation}
where $[\cdot]_k$ denotes the $k$-th element of a vector. After the optimal $\Theta_j$ is obtained, the quantiles $q_{ijt}^1,\dots,q_{ijt}^K$ can be calculated through $\Psi(x_{ijt}|\Theta_j)$ and a sorting operation is applied on them to ensure the monotony property, as described in Section \ref{sec:pre.quantile.regression}.
After that, a linear interpolation among these quantiles is applied to approximate the quantile function of $R_{ijt}$, i.e., the inverse of $F_{ijt}$. Thus $F_{ijt}$ is also approximated, for further use in the following steps of our method.

\subsection{Multi-name Dependence Modeling}
\label{sec:dependence modeling}

Default dependence among loans is crucial for pricing any credit portfolio. Joint defaults occur more frequently than in the independent case. To model it, we first handle the issue of mixed discrete and continuous variables with R-copula, then use Gaussian copula to model the dependence among all loans' repayment rates.

For a fixed $t$, suppose we have obtained every $F_{ijt}$ from the previous step. In the scenario of non-performing credit pricing, $F_{ijt}$ is discontinuous at $r=0$ and continuous otherwise. There is a jump from zero to a positive value at $r=0$, which means borrowers may repay zero amount with a positive probability. Hence, we apply the R-copula approach described in the section of preliminaries and obtain the standard uniformly distributed $U_{ijt}$ after transforming $R_{ijt}$. Next, the one-factor Gaussian copula is applied as usual:
\begin{align}
V_{ijt} & = G^{-1}(U_{ijt}), \label{eqn:from u to v} \\
V_{ijt} & = \beta_{jt} M_t + \sqrt{1-\beta_{jt}^2} Z_{ijt}, \label{eqn:one factor copula model}
\end{align}
where $G$ is the cumulative function of standard Gaussian. Now $V_{ijt}\sim N(0,1)$, and both the latent $M_t \sim N(0,1)$, $Z_{ijt} \sim N(0,1)$. The subscript $t$ in $M_t$ means all loans in the pool in a fixed time share the same common factor. 
A large negative value of $M_t$ will cause all $V_{ijt}$ tend to have negative values at time $t$, thus small $U_{ijt}$ and netagive $R_{ijt}$ are more possible.
The subscript $jt$ in the parameter $\beta_{jt}$ means the loans in the same category share the same sensitivity to the common factor. This is called the homogeneous credit portfolio assumption which is used for simplifying the model.

\subsection{Dependence Parameter Estimation}
\label{sec:dependence parameter}

$\beta_{jt}$ is a measure of the dependence among all loans in a category in a fixed time. A large $\beta_{jt}$ will induce more risk into the credit portfolio because borrowers tend to have lower repayment rates. In credit derivatives of corporate bonds, a large $\beta$ means joint defaults will happen more likely. Thus, a precise estimate of this parameter is crucial for the pricing and risk management of credit derivatives. In credit derivatives of corporate bonds, $\beta$ can be implied or recovered from the market quotes and the implied $\beta$ can be used to price new credit derivatives. However, very differently, large pools of small consumer loans are not traded in a market in our situation. We have only one choice: estimating $\beta$ from historical data. 

The difficulty may come from the high-dimensional nature of the problem, i.e., typically a pool may have thousands or even millions of small consumer loans. Fortunately, our model framework is succinct enough to allow us to do the estimation. Suppose we observed the realizations of $R_{ijt}$ as $r_{ijt}$ and the corresponding realizations of $U_{ijt}$ as $u_{ijt}$ after the R-copula transform, then the realizations $v_{ijt}$ of $V_{ijt}$ can be obtained by Equation (\ref{eqn:from u to v}). From Equation (\ref{eqn:one factor copula model}), conditional on $M_t=m_t$ in the fixed time $t$ and when $j$ is also fixed, all $V_{ijt}$ are independently and identically distributed (i.i.d.). This inspires us to apply the generalized method of moments (GMM)\cite{cragg1983more} to estimate unknown parameters.

Specifically, when fixing $t$ and $j$, the conditional first and second moments of $V_{ijt}$ are:
\begin{align}
E[V_{ijt} | M_t=m_t] & = \beta_{jt} m_t, \\
\text{Var}[V_{ijt} | M_t=m_t] & = 1-\beta_{jt}^2.
\end{align}
when $m_t$ is also unknown, the above two equations have two unknown parameters $m_t$ and $\beta_{jt}$ which need estimation. Noticed that $m_t$ is shared by every $j$-th category, separate estimation for different $j$ will yield inconsistent values of $m_t$. Thus, we combine these two equations for all $j=1,\dots,J$. Now $J+1$ unknown parameters are in $2J$ equations, making GMM applied successfully. Replacing the expectations with empirical sums, we have the final equations:
\begin{align}
\overline{v}_{jt} & := \frac{1}{N_{jt}} \sum_{i\in I_{jt}} v_{ijt} = \beta_{jt} m_t,\quad j=1,\dots,J, \\
\overline{\text{Var}}_{jt} & := \frac{1}{N_{jt}-1} \sum_{i\in I_{jt}} (v_{ijt}-\bar{v}_{jt})^2 = 1-\beta_{jt}^2,\quad  j=1,\dots,J.
\end{align}
There may be no solutions to the above equation system, so we solve the squared loss minimization instead:
\begin{equation}
\min_{m_t,\beta_{1t},\dots,\beta_{Jt}} \sum_{j=1}^{J} (\beta_{jt} m_t - \overline{v}_{jt})^2 + \sum_{j=1}^{J} (1-\beta_{jt}^2 - \overline{\text{Var}}_{jt} )^2.
\end{equation}
Here we have finished the estimation for unknown parameters at time $t$. This estimation is separate for different $t$, such that we do not need long historical data. Actually, the cross-sectional information is rich enough for us to conduct such an estimation. At last, we will do this estimation many times and take an average of the estimated parameters to reduce the inaccuracy caused by additional randomness incurred by $U'$ in the R-copula approach.

\subsection{Simulation for Portfolio}

Our aim is to have an estimation or prediction of the full distribution of the entire portfolio's repayment rate at every time $t$, as defined in Equation (\ref{eqn:definition of R_t}). So far, we have prepared two components that are sufficient to achieve this goal. In Section \ref{sec:marginal prediction}, we have approached the marginal distribution of $R_{ijt}$ ($j=1,\dots,J$ and $i\in I_{jt}$) by a neural network-based model's output.
In Section \ref{sec:dependence modeling} and \ref{sec:dependence parameter}, we modeled and estimated the dependence among $R_{ijt}$ ($j=1,\dots,J$ and $i\in I_{jt}$) through the copula approach. These two components are enough to determine the joint distribution of $R_{ijt}$, thus are enough to determine the distribution of $R_{t}$ too.

We are unable to derive the analytical form of the joint distribution of $R_{ijt}$. Thus simulation method is adopted, and the workflow is the inverse of the modeling system. At every time $t$, in every simulation step, we start from Equation (\ref{eqn:one factor copula model}) and simulate a realization from $M_t$ and a realization from $Z_{ijt}$ for every $j$ and $i$. Then a realization of $V_{ijt}$ for every $j$ and $i$ is obtained by Equation (\ref{eqn:one factor copula model}).
We then apply Equation (\ref{eqn:from u to v}) as well as the inverse of R-copula to obtain a realization of $R_{ijt}$, $\forall j=1,\dots,J$ and $i\in I_{jt}$. By Equation (\ref{eqn:definition of R_t}), a realization of $R_t$ is calculated. This process can be repeated many times for us to have an empirical estimation of the distribution of $R_t$. A pricing formula will take this empirical distribution as an input, which is not the focus of our paper anymore.

\section{EXPERIMENTS}

In this section, we comprehensively describe the experiments we conduct. First, we collect consumer loan data from one of the biggest online platforms all over the world for credit lending to consumers. Then the proposed architecture is implemented on the large amount of non-performing loan data. We do fine analysis on the performance of our method, and also discuss its merits and limitations.

\subsection{The Data}

At the beginning of each month in the whole year of 2019, we consider it as the day of inspection and randomly select about 360,000 overdue loans from the central data warehouse to form the portfolio. 
According to their overdue stages, we split them into $J=17$ categories named as M$_1$, M$_2$, $\dots$, M$_{12}$, M$_{13-15}$, M$_{16-18}$, M$_{19-21}$, M$_{22-24}$, and M$_{24+}$, where M$_i$ means that the days overdue are more than $i$ months but less than $i+1$ months. There is a small chance that one loan may appear in different days of inspection, but is in different stages. Naturally, when $i$ grows, the number of loans in M$_i$ will decrease rapidly. To ensure the same stability of our model for every category, we select balanced numbers of loans for all overdue stages. For example, the number of loans in M$_1$ at the date 2019-01-01 is about 10,000, and so is it for M$_i$, $i=2,\dots,36$. We merge some stages to form the final $J=17$ categories.

\begin{figure}[t]
  \centering
  \includegraphics[width=\linewidth]{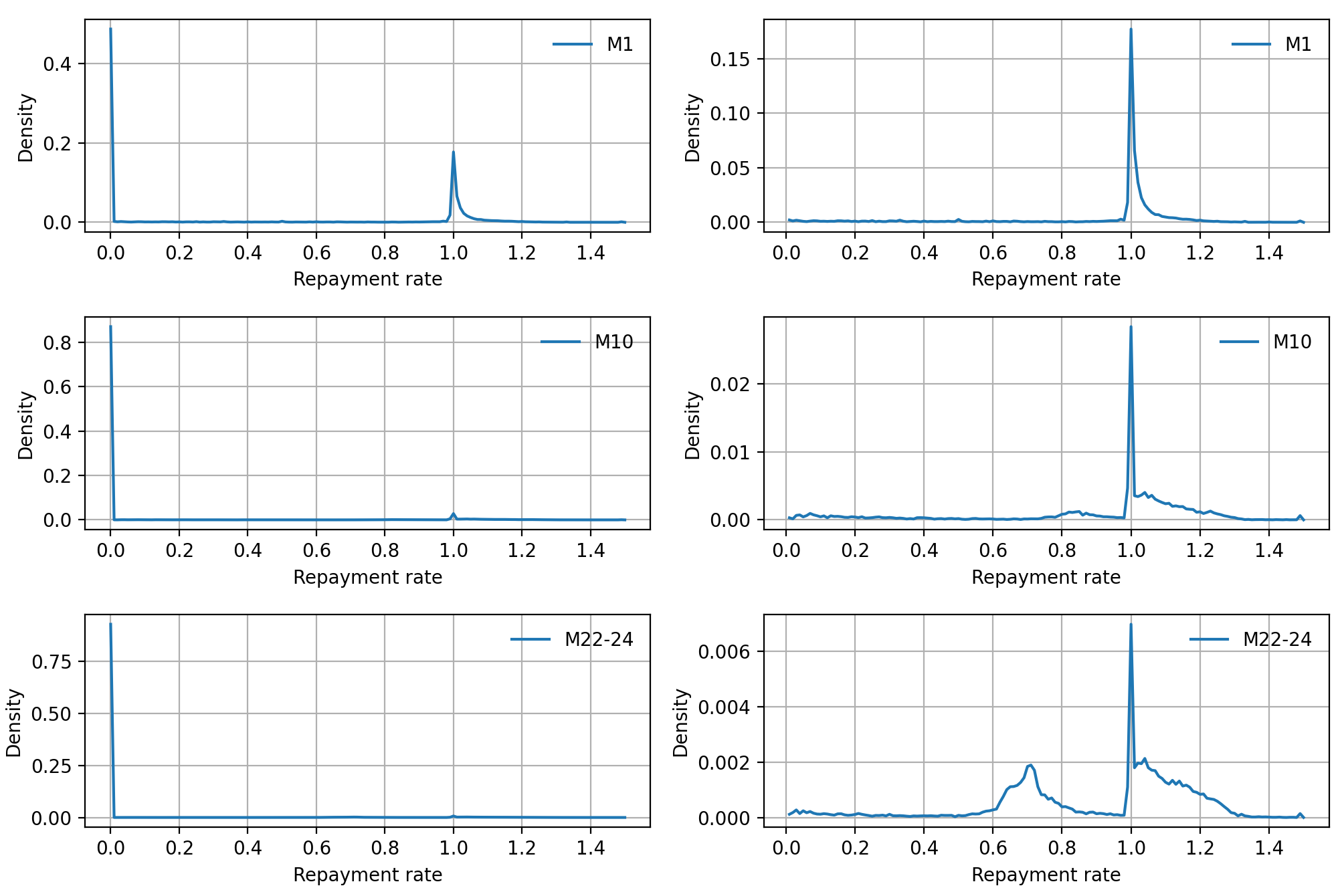}
  \vspace{-2em}
  \caption{The histogram of all loans' realized repayment rates in each category. The right column is the case when we delete zero repayment rates. For space limit, we only show the results of three categories.}
  \label{fig:unconditional histogram of repayment rates}
\end{figure}

With every loan, we extract its features at the time of inspection as well as its label which is the repayment rate in future one year. The features are 84-dimensional covering comprehensive information about the loan and the borrower, such as the loan's basic information, the borrower's repayment history, the borrower's searching and purchasing behaviors on the platform, and so on.
The repayment rate of each loan and the overall repayment rate of a group of loans are calculated by Equation \eqref{eqn:repayment rate single} and \eqref{eqn:definition of R_t}. The estimation and prediction of the latter is our final goal. Before starting, we first have a look at the histogram of all loans' realized repayment rates in each category, as shown in Figure \ref{fig:unconditional histogram of repayment rates}. We can be convinced by the figure that: 1) the single repayment rate is indeed a mixed variable of discrete and continuous; 2) as overdue stage increases, the probability of zero repayment raises significantly.

\begin{figure*}[t]
  \centering
  \includegraphics[width=\linewidth]{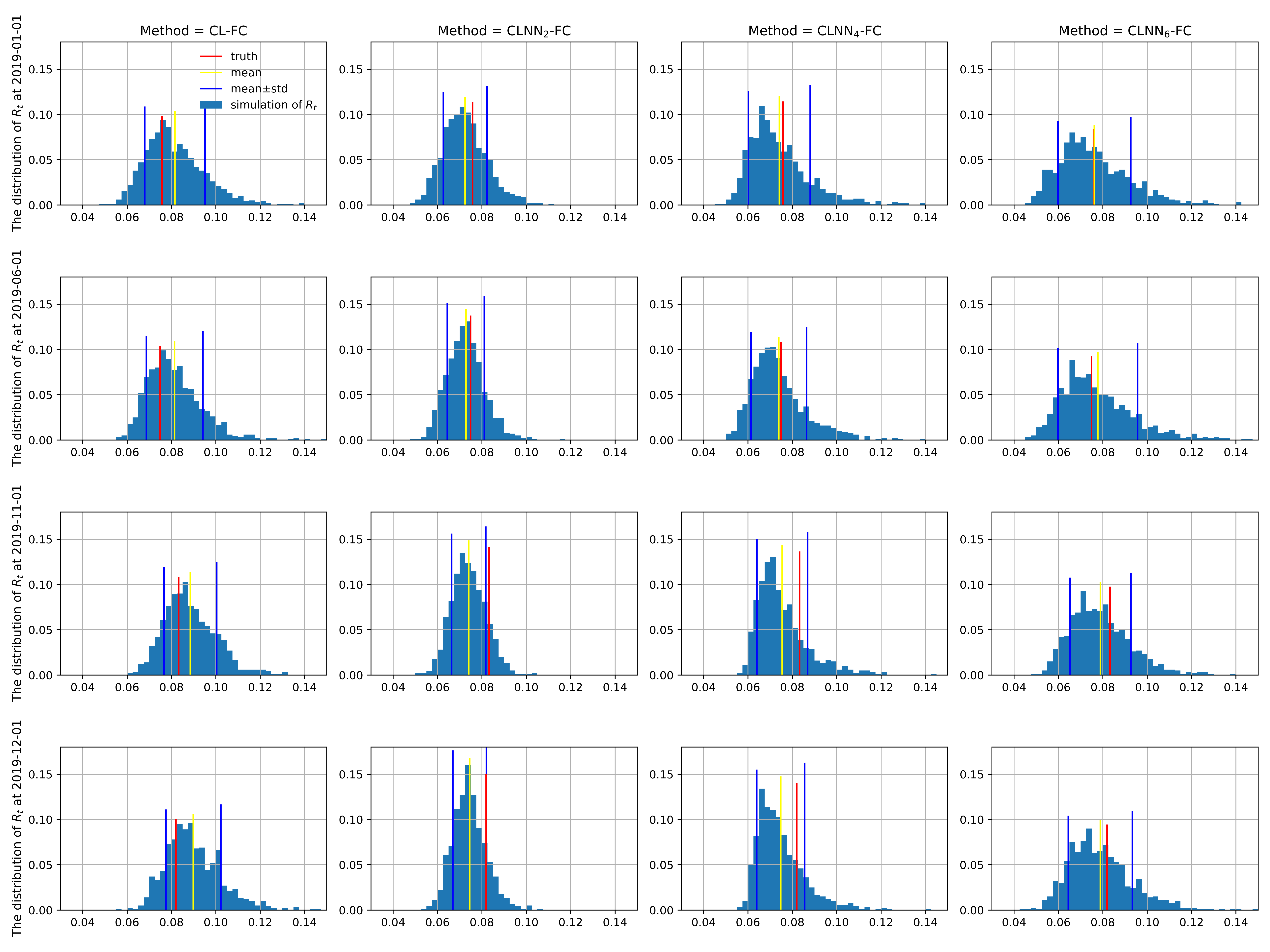}
  \vspace{-2em}
  \caption{The empirical distribution of $R_t$, or, the histogram of the simulation of $R_t$.
The yellow line and dark blue lines are the mean and mean$\pm$std of the simulation respectively. The red line is the true observed value of $R_t$. Here we only show the results of the dates 2019-01-01, 2019-06-01, 2019-11-01, and 2019-12-01 from top to down.
 Each column represents one version of our method: CL-FC, CLNN$_2$-FC, CLNN$_4$-FC, and CLNN$_6$-FC.}
\label{fig:histograms}
\end{figure*}
\begin{table*}[t]
  \caption{Related statistics of the predicted distribution of $R_t$.}
  \vspace{-1em}
    \label{tab:distribution statistics}
    \begin{tabular}{c|ccccc|ccccc}
    \hline
              &        &        &CL-FC   &          &          &        &        &CLNN$_2$-FC&          &      \\
    Date      & Truth  &Mean    &Std     & Skewness & Kurtosis & Truth  &  Mean  &  Std   & Skewness & Kurtosis\\
    \hline 2019-01-01  & 0.0758 & 0.0815 & 0.0136 & 0.8852   & 1.4365   & 0.0758 & 0.0725 & 0.0099 & 0.5158   & 0.3759  \\
    2019-06-01  & 0.0749 & 0.0814 & 0.0127 & 1.2853   & 3.8079   & 0.0749 & 0.0727 & 0.0083 & 0.4934   & 0.9094  \\
    2019-11-01 & 0.0832 & 0.0885 & 0.0102 & 0.6639   & 0.8275   & 0.0832 & 0.0740 & 0.0077 & 0.2699   & 0.1589  \\
    2019-12-01 & 0.0819 & 0.0898 & 0.0107 & 0.6926   & 0.9156   & 0.0819 & 0.0744 & 0.0076 & 0.5126   & 0.6511  \\
    \hline    &        &     &CLNN$_4$-FC&          &          &        &        &CLNN$_6$-FC&       &         \\
    Date      & Truth  &Mean &Std        & Skewness &Kurtosis  &Truth   &Mean    &Std     & Skewness & Kurtosis\\
    \hline 2019-01-01  & 0.0758 & 0.0741 & 0.0139 & 1.2779   & 2.2788   & 0.0758 & 0.0762 & 0.0164 &1.0864    & 2.1673  \\
    2019-06-01  & 0.0749 & 0.0738 & 0.0125 & 1.1774   & 2.0358   & 0.0749 & 0.0778 & 0.0179 & 1.7292   & 6.3886  \\
    2019-11-01 & 0.0832 & 0.0754 & 0.0114 & 1.4677   & 2.9049   & 0.0832 & 0.0789 & 0.0137 & 0.8954   & 1.4838  \\
    2019-12-01 & 0.0819 & 0.0747 & 0.0108 & 1.4480   & 3.2345   & 0.0819 & 0.0789 & 0.0144 & 0.9268   & 1.6873  \\
    \hline
    \end{tabular}
\end{table*}

\subsection{Settings}

We first use the data collected from the date 2019-01-01 to 2019-10-01 for learning and analysis. The data in this period is randomly split into two parts for model training and testing respectively. Thus we have 10 time points and at each, a collection of $J=17$ categories of loans is in the training set and another collection is in the testing set. We leave the data in the period from 2019-11-01 to 2019-12-01 as another testing set so that it has no overlap with the training set in time. It should be noted that the parameters $\beta_{jt}$ are not estimated for these two months. Hence we use the average of the previous estimations of $\beta_{jt}$ for every $j$, as the prediction for the new $\beta_{jt}$.

In the quantile regression part of our method, 99 probability levels regularly spreading over $(0,1)$ are chosen for approximating the marginal distribution of each loan's repayment rate through the quantiles. More specifically, $[\tau_1,\tau_2,\dots,\tau_K]=[0.01,0.02,\dots,0.99]$ is chosen in Equation \eqref{eqn:quantile regression function chosen}. Also in Equation \eqref{eqn:choice of psi}, the number of neural network layers $L$ is chosen to be 2, 4, or 6 for checking the performance difference. We denote our whole model with $L$ as CLNN$_L$-FC, where CL is the abbreviation for clip and linear operations, NN is for neural network, and FC is for factor copula. Besides, $L$ can also be 0, which means the features are input into the linear and clip operations directly. We denote such a version of our model as CL-FC and implement it for testing the performance.

For the neural networks, we select the sigmoid function as the activation function in all hidden layers. For CLNN$_2$-FC, the hidden layer sizes are $[128,64]$, and for CLNN$_4$-FC and CLNN$_6$-FC, the hidden layer sizes are $[128, 64, 32, 16]$ and $[64, 64, 32, 32, 16, 8]$ respectively.
The linear layer has 99 output nodes and the output is clipped to the range $[0,1.5]$ finally. The repayment rate may be larger than 1 because of the penalty interest.
We use Adam \cite{DBLP:journals/corr/KingmaB14} as the optimization algorithm and the learning rate is 0.001 by default. The number of iteration steps is 1000. We adopt these hyper-parameters without further adjustment, because in our experience there is no big difference when choosing different hyper-parameters.
Then, following the workflow of our method, the parameters of the dependence model are estimated. To reduce the error caused by extra randomness in R-copula, this estimation is performed ten times and the average estimated parameters are taken. At last, in the simulation for portfolio loss distribution, 1000 realizations of ${R_t}$ are obtained to approximate the distribution of $R_{t}$.

What we propose here is a bottom-up model, i.e., dealing with each individual loan first, followed by combining them to get the portfolio loss distribution. Another class of methods for credit portfolio modeling is called top-down, i.e., the portfolio loss distribution is specified directly without using the information of individual constituent loans \cite{giesecke2009portfolio}. To confirm the necessity of constructing a bottom-up model as we do, and also to demonstrate the merits of our model, we implement a benchmark top-down model for comparison which simply uses past four months' moving average of each category's overall repayment rates as the prediction of next repayment rate for that category. The prediction of the whole portfolio's repayment rate is done by combining all categories together. However, this method cannot obtain the distribution of $R_t$.

\subsection{W-test}

Through the simulation, we can predict the distribution of $R_t$, which is denoted as $F_t$.
Ideally, the variable $F_t(R_t)$ should follow a uniform distribution over $(0,1)$.
So, if we have predicted $F_t$ for $R_t$, $t\in \mathcal{T}$, and if we also have the observed realization of $R_t$ which is denoted as $r_t$, we can test if the quantities $F_t (r_t)$, $t\in \mathcal{T}$ follow a uniform distribution over $(0,1)$.
Furthermore, this test is equivalent to testing if $G^{-1}(F_t(r_t))$ follow a standard normal distribution, where $G$ is the cumulative function of standard normal.
Therefore, finally we will do a normality test on $G^{-1}(F_t(r_t))$, $t\in \mathcal{T}$.

The number of our observation dates is 10 in the first testing test. So, Shapiro-Wilk test (W-test) can be used to test if a small set of samples follow the normal distribution \cite{shapiro1965analysis}. 
Suppose the data samples we need to test are $y_t = G^{-1}(F_t(r_t))$, and $y_{(1)} \leq y_{(2)} \leq \cdots \leq y_{(n)}$ are the order statistics of $y_t$, $t\in \mathcal{T}$ (we denote $n=|\mathcal{T}|$ here),
then the statistics of W-test is defined as:
\begin{equation}
    W = \dfrac{\left(\sum_{i=1}^{n}{a_iy_{(i)}}\right)^2}{\sum_{i=1}^{n}({y_{(i)}-\bar{y}})^2},
\end{equation}

where the coefficients $a_1, a_2, \cdots, a_n$ have specific values when the sample size is $n$.
The null hypothesis of the test is that the data follows the normal distribution. 
This hypothesis is rejected if $W$ is below a critical value under a certain confidence level, say 95\% chosen by us in this paper. 
A $p$-value close to 1 is an indicator of good performance, and a $p$-value less than 0.05 is a strong indicator of rejection of the null hypothesis.

\subsection{Results}


We first examine the predicted distribution of $R_t$ obtained by our method on the testing set, through plotting the histogram of the simulated realizations of $R_t$. Figure \ref{fig:histograms} shows the histograms given by the four versions of our method: CL-FC, CLNN$_2$-FC, CLNN$_4$-FC, and CLNN$_6$-FC. For the limit of pages, we only place the histograms of the portfolios at four dates: 2019-01-01, 2019-06-01, 2019-11-01, and 2019-12-01. The latter two have no overlap with the dates of the training set. In Figure \ref{fig:histograms}, the mean and mean$\pm$std of the simulation of each histogram, as well as the true observed repayment rate of $R_t$, are also plotted. We can see that in many cases, the true observation lies in an appropriate range of the histogram. We also list the higher-order moments of every predicted distribution of $R_t$ in Table \ref{tab:distribution statistics}, which may be the extra inputs of a pricing formula.

To rigorously evaluate the performance of the distribution prediction, we conduct W-test for each version of our method over the first testing set which has 10 dates (W-test requires the dataset size $n\ge 8$). All test results are reported in Table \ref{tab:wtest}.
As shown in Table \ref{tab:wtest}, only linear regression-based method gets a significant rejection of the hypothesis that $G^{-1}(F_t(r_t))$, $t\in\mathcal{T}$ follow a standard Gaussian. In some sense, the performances of CLNN$_2$-FC and CLNN$_4$-FC are good enough because the W-test statistics are close to 1. Thus a mild layer size is enough in the specific design of our architecture. This may be because that the features we select as the input of the quantile regresison have been proven previously to contain the most information relevant to the default prediction.
\begin{table}[t]
\caption{The W-test results. The CL-FC method yields a significant rejection. CLNN$_2$-FC and CLNN$_4$-FC produce W-test statistics that are close to 1.}
\vspace{-1em}
  \label{tab:wtest}
  \begin{tabular}{ccc}
    \toprule
    Method & W-test statistics & $p$-value\\
    \midrule
    CL-FC & 0.7916 & 0.0114\\
    CLNN$_2$-FC & 0.9759 & 0.9397\\
    CLNN$_4$-FC & 0.9798 & 0.9641\\
    CLNN$_6$-FC & 0.9602 & 0.7886\\
  \bottomrule
\end{tabular}
\end{table}

To further evaluate the performance of our method in a different way, we choose the mean absolute percentage error (MAPE) as another  evaluation criteria. If we use $\hat{r}_t$ to denote the mean of the predicted distribution of $R_t$ (the yellow lines in Figure \ref{fig:histograms}), then the criteria MAPE can be calculated as:
\begin{equation}
  \text{MAPE} = \dfrac{1}{|\mathcal{T}|}\sum_{t\in \mathcal{T}}\dfrac{|r_t-\hat{r}_t|}{ |r_t|}.
\end{equation}
In MAPE, we take $\hat{r}_t$ as the predicted value of $r_t$, and it is consistent with the idea that the predicted value should be as close as possible to the true observed value $r_t$. 
Since we have two testing sets, we report the results on both sets for comprehensive comparisons.
We also include the performance of the top-down method we implement into comparisons.

Table \ref{tab:mape} presents the MAPE values of different methods on the first testing set covering 10 months. 
Not surprisingly, the neural network-based methods work much better than linear regression-based method.
Among the neural network-based methods, CLNN$_4$-FC yields the best performance. More crucially, the top-down method performs much worse than our methods do, indicating the absolute necessity of constructing a bottom-up system for the pricing of non-performing consumer credit portfolios, as we do in this paper.
\begin{table}[t]
  \caption{The MAPE results. CLNN$_4$-FC gives the best performance while the top-down method is the worst.}
  \vspace{-1em}
  \label{tab:mape}
  \begin{tabular}{ccc}
    \toprule
    Method&MAPE\\
    \midrule
    CL-FC & 9.36\%  \\
    CLNN$_2$-FC & 4.10\% \\
    CLNN$_4$-FC & 3.38\% \\
    CLNN$_6$-FC & 4.20\%\\
    top-down& 10.81\%\\
  \bottomrule
\end{tabular}
\end{table}

Table \ref{tab:mape(outofsample)} presents the MAPE results over the second testing set covering the non-overlap two months. Different from the first MAPE table, CLNN$_6$-FC works best in the non-overlap case. Nevertheless, the significance of this phenomenon may be restricted to the sample size. We could not have more data points over time because constructing such a dataset needs a observation period of future one year. So we should have records of the loans up to the end of the year 2020, which is unavailable when this research is done.
\begin{table}[t]
  \caption{The MAPE results over the non-overlapping testing set. CLNN$_6$-FC gives the best performance while the top-down method is the worst.}
  \vspace{-1em}
  \label{tab:mape(outofsample)}
  \begin{tabular}{ccc}
    \toprule
    Method & MAPE (non-overlap)\\
    \midrule
    CL-FC & 8.00\%\\
    CLNN$_2$-FC & 10.10\%\\
    CLNN$_4$-FC & 9.09\%\\
    CLNN$_6$-FC & 4.42\%\\
    top-down& 14.09\% \\
  \bottomrule
\end{tabular}
\end{table}

To conclude all the experiment results, the method we construct is able to capture both the marginal distribution of every loan and the dependence structure among loans in a portfolio, and consequently can yield good prediction of the distribution of overall repayment rate of a credit portfolio containing non-performing consumer loans. Thus this workflow or architecture has great potential of being successfully applied to the real business domain.

\subsection{Drawbacks and Future Works}

As one can find, the methods or techniques we adopt here are mainly traditional statistical ones, except that in individual loan's repayment forecasting we use neural networks. This is primarily because AI currently lacks (popular) methodologies for: i) effective probabilistic forecasting (as opposed to point forecasting in supervised machine learning), which is of great need in finance; ii) the dependence modeling of extremely high-dimensional variables. For the second one, \cite{yan2019cross} recently proposed a generative machine learning model for capturing tail dependence structure among stock returns. However, it remains a quite challenging task and needs more further research works.

Another potential drawback of our approach is the homogeneous assumption in the dependence modeling, i.e., the constant $\beta$ across a category of loans. Although it is a common setting in credit portfolio pricing frameworks for simplifying, it may miss the possible more complex dependence structure in the data. Future works can aim at alleviating this drawback in a more data-driven manner.

\section{Conclusions}

We have described in detail our bottom-up system for the overdue loan portfolios' repayment risk prediction. Every part of our architecture is elaborately designed and finally, the experiments demonstrate the success of our system. It can be used in the pricing and risk management of non-performing consumer credit portfolios and may benefit both the new-emerging risk transferring business and the ordinary risk management tasks.

In this paper, we do not explicitly give the price of a credit portfolio. What we offer is the expected profit or cash flow (the mean repayment) generated by holding the portfolio, as well as the uncertainty measures such as volatility (standard deviation), skewness, and kurtosis. One can then determine his/her own acceptable price based on the basic finance principle of risk-return tradeoff, i.e., willing to pay the lower price when the profit uncertainty raises. In other words, the expected return should increase as the risk raises. The specific price should depend on personal risk aversion level or pricing kernel.

Future works may include incorporating macroeconomic risk drivers into the system and studying tail dependencies among loans.
The macroeconomic cycle may have effects on the repayment correlation structure of overdue loans. Besides, whether the repayment correlations are heterogeneous for different regions and industries is an interesting question. Finally, analogous to corporate bonds and CDO pricing problem, whether the repayments of overdue loans exhibit tail dependencies such that extreme bad events happen unexpectedly.
All of these are worthy of further in-depth studies.

\bibliographystyle{ACM-Reference-Format}
\bibliography{ICAIF_paper_38}

%
%
%
%

\end{document}